\PassOptionsToPackage{pdftex}{graphicx}
\documentclass[shortnote,twocolumn]{jpsj3}
\usepackage{graphicx,bm,amssymb,amsmath,url}
\usepackage{tikz}
\usepackage[colorlinks=true,allcolors=blue]{hyperref}

\makeatletter
\let\balancecolumns\relax
\let\balancing@outputdblcol\relax
\let\balancing@outputpage\relax
\providecommand\ext@figure{lof}
\providecommand\ext@table{lot}
\def\@lastpagebalancing{false}
\makeatother

\title{Finite-Basis Duality Estimate for the Surface-Code Threshold under Correlated Bit-Flip Errors}

\author{
    Masayuki \textsc{Ohzeki}\thanks{E-mail: mohzeki@tohoku.ac.jp}
}

\inst{
    Graduate School of Information Sciences, Tohoku University, Sendai 980-8579, Japan\\
    Department of Physics, Institute of Science Tokyo, Meguro, Tokyo 152-8551, Japan\\
    Research and Education Institute for Semiconductors and Informatics, Kumamoto University, Kumamoto 860-8555, Japan\\
    Sigma-i Co., Ltd., Minato, Tokyo 108-0075, Japan
}

\abst{
We apply finite-basis duality to a statistical-mechanical model introduced by Chubb and Flammia for the surface code under spatially correlated bit-flip noise.  Their mapping gives a random-bond Ising model with both two-body edge interactions and four-body face interactions.  
The single-equation estimate based on the duality analysis is slightly deviated from their Monte Carlo estimate \(p_c=0.1004(6)\).  Following the finite-basis and graph-polynomial strategy, we instead use periodic and twisted-periodic sectors on toroidal bases. 
We have obtained an improved estimate of the critical point, $p_c = 0.10348$.
}

\kword{quantum error correcting code, surface code, correlated noise, duality, threshold}

\begin{document}
\makeatletter
\let\balancecolumns\relax
\let\balancing@outputdblcol\@outputdblcol
\makeatother
\maketitle

A key challenge in random spin systems is determining critical points.  Among various analytical tools, duality analysis stands out as an unexpectedly powerful and conceptually simple method.  A classic example is the Kramers--Wannier duality~\cite{Kramers1941}.  
Even for disordered systems, in conjunction with the  replica method, it led to accurate conjectures~\cite{Nishimori1979,Nishimori2002,Maillard2003,Nishimori2006}.
These ideas have also been used to estimate thresholds of quantum error-correcting codes and related random planar-lattice models~\cite{Dennis2002,Ohzeki2009color,Ohzeki2012,Ohzeki2012duality,Bombin2012}.  
A further refinement combined real-space-renormalized duality with real-space renormalization and graph-polynomial ideas~\cite{Ohzeki2008hl, Ohzeki2009regular, Ohzeki2015}.  
Recent extensions include symmetric-group duality for the random tensor network and a minimal duality estimate for nearest-neighbor correlated-error surface codes~\cite{Ohzeki2024,Ohzeki2026minimal}.

Chubb and Flammia generalized a statistical-mechanical mapping to stabilizer and subsystem codes under correlated Pauli noise~\cite{Chubb2021}.  
Indeed, they studied the surface code under mildly correlated bit-flip noise and used Monte Carlo simulations of the associated statistical-mechanical model to estimate the threshold as \(p_{\rm corr}=10.04(6)\%\).  
The purpose of this short note is to show that the correlated-noise threshold can be estimated, with good accuracy, by finite-basis duality again.

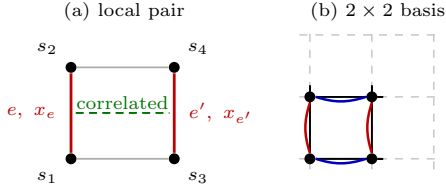
\begin{figure}[t]
\centering
\begin{tikzpicture}[
  scale=0.78,
  spin/.style={circle,fill=black,inner sep=1.45pt},
  edge/.style={line width=0.65pt},
  pair/.style={red!75!black,line width=1.05pt},
  four/.style={green!45!black,line width=0.85pt,densely dashed},
  hp/.style={red!75!black,line width=0.95pt,bend left=18},
  vp/.style={blue!75!black,line width=0.95pt,bend right=18},
  seam/.style={dashed,line width=0.55pt}
]
  \node at (0.9,2.50) {\scriptsize (a) local pair};
  \node[spin,label=below left:{\scriptsize \(s_1\)}] (a1) at (0,0) {};
  \node[spin,label=above left:{\scriptsize \(s_2\)}] (a2) at (0,1.55) {};
  \node[spin,label=below right:{\scriptsize \(s_3\)}] (a3) at (1.75,0) {};
  \node[spin,label=above right:{\scriptsize \(s_4\)}] (a4) at (1.75,1.55) {};
  \draw[edge,gray!65] (a2)--(a4);
  \draw[edge,gray!65] (a1)--(a3);
  \draw[pair] (a1)--node[left=4pt,fill=white,inner sep=1pt]{\scriptsize \(e,\ x_e\)} (a2);
  \draw[pair] (a3)--node[right=4pt,fill=white,inner sep=1pt]{\scriptsize \(e',\ x_{e'}\)} (a4);
  \draw[four] (0.08,0.78)--node[above,fill=white,inner sep=1pt]{\scriptsize correlated} (1.67,0.78);

  \begin{scope}[xshift=4.05cm]
    \node at (1.15,2.50) {\scriptsize (b) \(2\times2\) basis};
    \foreach \x in {0,1,2}{
      \draw[seam,gray!45] (\x*1.05,-0.18) -- (\x*1.05,2.22);
      \draw[seam,gray!45] (-0.18,\x*1.05) -- (2.22,\x*1.05);
    }
    \foreach \x in {0,1}{
      \foreach \y in {0,1}{
        \node[spin] (t\x\y) at (\x*1.05,\y*1.05) {};
      }
    }
    \draw[edge] (t00)--(t10);
    \draw[edge] (t01)--(t11);
    \draw[edge] (t00)--(t01);
    \draw[edge] (t10)--(t11);
    \draw[edge] (t10) to[out=0,in=0] (t00);
    \draw[edge] (t11) to[out=0,in=0] (t01);
    \draw[edge] (t01) to[out=90,in=90] (t00);
    \draw[edge] (t11) to[out=90,in=90] (t10);
    \draw[hp] (0,0.10) to (0,0.95);
    \draw[hp] (1.05,0.10) to (1.05,0.95);
    \draw[vp] (0.10,0) to (0.95,0);
    \draw[vp] (0.10,1.05) to (0.95,1.05);
  \end{scope}
\end{tikzpicture}
\caption{(Color online) (a) Local correlated pair used by Chubb and Flammia.  Opposite plaquette edges \(e,e'\) carry quenched variables \(x_e,x_{e'}\), while the statistical-mechanical Ising spins \(s_v\) live on vertices.  The pair gives two edge terms and a four-spin term.  (b) Smallest toroidal finite basis used in Eq.~\eqref{eq:twist}; red and blue arcs denote non-overlapping correlated pairs, and periodic or antiperiodic seams give the four sectors \(Z_{\alpha\beta}\).}
\label{fig:cluster}
\end{figure}

Let \(\mathcal E_0=I\) and \(\mathcal E_1=X\) as flip error, and define the error model as in Fig.~\ref{fig:cluster}(a)
\[
P_{ab}=\Pr(E_e=\mathcal E_a,E_{e'}=\mathcal E_b),
\]
where \(a,b=0,1\) label the error types on \(e,e'\).  
Then
\begin{align}
P_{00}&=(1-p)(1-p_-),&
P_{01}&=p(1-p_+),
\nonumber\\
P_{10}&=(1-p)p_-, & 
P_{11}&=p p_+,
\label{eq:pairprob}
\end{align} 
and 
\begin{equation}
p_-=\frac{p}{1-p+\eta p} \quad p_+=\eta p_-.
\label{eq:ppm}
\end{equation}
We define $\eta=p_+/p_-$.
Only \(p\) and \(\eta\) are independent parameters.

For \(a,b=0,1\), set \(x_a=(-1)^a\) and \(x_b=(-1)^b\). 
We expand the local probability as
\begin{equation}
\log P_{ab}=C+K_1x_a+K_2x_b+K_{12}x_ax_b ,
\end{equation}
or, equivalently,
\[
(C,K_1,K_2,K_{12})
=\frac14\sum_{a,b=0}^1
(1,x_a,x_b,x_ax_b)\log P_{ab}.
\]
The local Boltzmann factor is then
\begin{equation}
W_{ab}(r_1,r_2)=
\exp\{K_1x_ar_1+K_2x_br_2+K_{12}x_ax_br_1r_2\},
\label{eq:localweight}
\end{equation}
where \(r_i=\prod_{v\in\partial e_i}s_v\) with \(s_v=\pm1\).  
The \(K_{12}\) term is the four-spin interaction.  
For \(K_1=K_2\equiv J_2\) and \(K_{12}\equiv J_4\), the corresponding statistical-mechanical model satisfying the Nishimori-condition is
\begin{equation}
H_E=
-J_2\sum_e x_e\prod_{v\in\partial e}s_v
-J_4\sum_{e\sim e'}x_ex_{e'}
\prod_{v\in\partial(ee')}s_v .
\label{eq:ham}
\end{equation}
Here \(\partial e\) is the two endpoints of \(e\), \(e\sim e'\) denotes the chosen opposite-edge pair across one plaquette, and \(\partial(ee')=\partial e\,\triangle\,\partial e'\), which is the set of four plaquette vertices for such a pair.

We now define the principal factors used in the duality estimate.  
Following the replica method, we obtain the \(n\)-copy local weight by taking the average over the quenched pair-error state \(ab\).  The principal factor \(x_0(n)\) is the component in which all replicas are in the trivial relative-spin configuration, \(r_1^{(\alpha)}=r_2^{(\alpha)}=1\) for \(\alpha=1,\ldots,n\).  
The dual principal factor \(x_0^\ast(n)\) is the corresponding zero-Fourier component after applying the \(Z_2\) duality transform to the relative-spin variables in each replica.  
\begin{align}
x_0(n)&=\sum_{a,b=0}^1 P_{ab}W_{ab}(1,1)^n,\\
x_0^\ast(n)&=2^{-n}\sum_{a,b=0}^1P_{ab}
\left\{\sum_{r_1,r_2=\pm1}W_{ab}(r_1,r_2)\right\}^n .
\end{align}
The factor \(2^{-n}\) is the product of the two Kramers--Wannier normalizations \(2^{-n/2}\), one for each edge in the pair.
The sums are therefore explicit finite sums over the four pair-error configurations \(ab=00,01,10,11\) and the four relative-spin states \(r_1,r_2=\pm1\).  
Once \(x_0(n)\) and \(x_0^\ast(n)\) are obtained, the single-cell duality estimate imposes $x_0(n)=x_0^\ast(n) $ and then takes the quenched limit \(n\to0\).  
We obtain
\begin{equation}
-\sum_{a,b=0,1}P_{ab}\log_2P_{ab}=1 .
\label{eq:singlepair}
\end{equation}
For \(\eta=1\), we reproduce the usual single-bond estimate \(p_c=0.11003\ldots\).  
For the positively correlated case \(\eta=2\), it gives $
p_c^{\rm pair}=0.11105\ldots$. 
In contrast, the Monte Carlo calculation gives \(0.1004(6)\) lower than the independent-noise Monte Carlo value \(0.10917\) ~\cite{Chubb2021}.
Thus positive spatial correlation makes decoding harder, while the single-pair principal-factor estimate predicts the opposite trend.  

We therefore use the finite-basis duality idea of Ref.~\cite{Ohzeki2015}. 
We first take a small size of clusters to manipulate the real-space renormalization.
For a basis \(B\) closed into a torus, we define \(Z_{\alpha\beta}\) with periodic or antiperiodic boundary conditions \(\alpha,\beta=\pm\).  The estimate is
\begin{equation}
\left[\log\{Z_{++}+Z_{+-}+Z_{-+}+Z_{--}\}\right]
-[\log Z_{++}]=\log2 ,
\label{eq:twist}
\end{equation}
where \([\cdots]\) denotes the quenched average.   
This is the finite-basis replacement of the single-pair principal-factor equation: the local quantities \(x_0,x_0^\ast\) are replaced by the trivial and twisted torus-sector partition functions.

The smallest nontrivial basis used here is shown in Fig.~\ref{fig:cluster}(b).  The correlated pair factors are placed as a non-overlapping matching on a torus.  This choice has the important check that, when \(\eta=1\), the \(2\times2\) calculation reduces to the ordinary square-lattice RBIM finite-basis estimate:
\begin{equation}
p_c^{L=2}(\eta=1)=0.10863\ldots .
\end{equation}
For the correlated case \(\eta=2\), exact enumeration of the disorder on the same basis gives
\begin{equation}
p_c^{L=2}(\eta=2)=0.10120\ldots .
\label{eq:result}
\end{equation}
This is already very close to the Monte Carlo value \(0.1004(6)\) reported in Ref.~\cite{Chubb2021}. To test the systematic improvement strategy, we also evaluated a \(4\times4\) toroidal basis.  
On this basis there are 16 correlated opposite-edge pairs, hence 32 individual edge variables \(x_e\).  
A disorder configuration \(E\) assigns \(I\) or \(X\) to each of these 32 edges, with probability given by the product of the 16 pair probabilities \(P_{00},P_{01},P_{10},P_{11}\).  Let
\[
m(E)=\#\{e:x_e=-1\}
\]
be the number of bit-flip edges in \(E\).  
The truncated quenched average with cutoff \(k\) is obtained by summing exactly only those configurations with \(m(E)\le k\):
\[
[F(E)]_{k}
=
\frac{1}{\rho_k}
\sum_{E:m(E)\le k}P(E)F(E),
\qquad
\rho_k=\sum_{E:m(E)\le k}P(E).
\]
It is the maximum number of flipped edge variables retained in the deterministic disorder sum.  For \(k=12\), the calculation includes all configurations with up to 12 \(X\) edges among the 32 edge variables.  
Near the root this retains a probability weight \(\rho_{12}\simeq0.999975\) and involves \(462,411,533\) disorder configurations.  
The deterministic result is
\begin{equation}
p_c^{L=4,k=12}(\eta=2)=0.10348\ldots .
\label{eq:l4result}
\end{equation}
It is the root of Eq.~\eqref{eq:twist} with the truncated quenched average.  The movement from the single-pair result \(0.11105\ldots\) to the finite-basis values shows that the geometry of correlated frustration and domain-wall insertion is essential.

\begin{table}[t]
\centering
\caption{Threshold estimates for the Chubb--Flammia correlated bit-flip model with \(\eta=2\).}
\label{tab:results}
\begin{tabular}{lc}
\hline
method & \(p_c\) \\
\hline
single correlated pair & 0.11105 \\
\(2\times2\) finite basis, exact & 0.10120 \\
\(4\times4\) finite basis, \(k=12\) & 0.10348 \\
Monte Carlo~\cite{Chubb2021} & 0.1004(6) \\
\hline
\end{tabular}
\end{table}

The present calculation demonstrates that finite-basis duality is useful beyond independent-noise RBIMs and remains effective with explicit four-spin interactions generated by spatially correlated errors.  Since our method uses local Boltzmann weights, quenched averages, and twisted-sector partition functions rather than a full Monte Carlo phase diagram, it gives a practical way to screen quantum-error-correcting-code thresholds over broad correlated-noise families.

\begin{acknowledgment}
We received financial support from the Cross-ministerial Strategic Innovation Promotion Program (SIP) of the Cabinet Office (No. 23836436).
\end{acknowledgment}

\end{document}